# The Role of Spreadsheets in the Allied Irish Bank / Allfirst Currency Trading Fraud

Raymond J. Butler
ray.butler@virgin.net
CAS Salford

*This brief paper outlines how spreadsheets were used as one of the vehicles for John Rusnak's fraud and the revenue control lessons this case gives us.*

## 1. INTRODUCTION

The interim report [AIB Group, 2002] into the fraud(s) perpetuated by the "Rogue Trader" John Rusnak against the AIB subsidiary Allfirst Bank reports that:

> *"Mr. Rusnak circumvented the controls that were intended to prevent any such fraud by manipulating the weak control environment in Allfirst's treasury; notably, he found ways of circumventing changes in control procedures throughout the period of his fraud."*

This brief paper outlines how spreadsheets were used as one of the vehicles for Rusnak's fraud and the revenue control lessons this case gives us. The quotes are from AIB's interim report, the emphasis is the authors.

## 2. HOW DID RUSNAK CIRCUMVENT THE CONTROLS ?

One of the several ways in which Rusnak circumvented controls was by manipulating spreadsheet models used by the bank's internal control staff
.
2.1. Value at Risk

AIB used an internal Value at Risk (VaR) model (Appendix A) to determine potential losses. This model was manipulated by Rusnak:

> *"Rusnak manipulated the principal measure used by Allfirst and AIB to monitor his trading: Value at Risk (VaR). ..... by directly manipulating the inputs into the calculation of the VaR that were used by an employee in Allfirst's risk-control group. Thus, while that employee was supposed to independently check the VaR, she relied on a spreadsheet that obtained information from Rusnak's personal computer and that included figures for so-called 'holdover' transactions—entered into after a certain hour toward the end of each day. But these transactions were not real and.... were not even entered on to the bank's trading software. They were simply a way to manipulate the spreadsheet used to calculate the VaR. A simple check to see if the holdover figures were captured in the next day's trading activity would have caught this scheme."*

## 2.2. Foreign Exchange Rates

Since 1995 concerns had been raised that monthly foreign exchange prices were not obtained independently of the traders. Sporadic attempts were made obtain independent pricing but these soon lapsed. In 2000, a risk assessment analyst responsible for treasury raised the concern that :

> *".. daily rates were not being obtained from an independent source. At one time, revaluation rates were printed from Reuters and checked by treasury risk control. But in 2000, the treasury risk control analyst proposed developing a spreadsheet that would accomplish a direct data drop from Reuters and eliminate the need to manually check rates."*

Rates were downloaded from Rusnak's Reuters terminal to his personal computer's hard disk drive, and then fed into a database which was accessible to the front, back, and middle offices. The risk assessment analyst learned of this foreign exchange pricing spreadsheet. Her notes state:

> *"This is a failed procedure"*

and

> *"technically, the trader/s could manipulate the rates."*

She asked the risk control analyst why operations could not obtain the rates independently and he replied that Allfirst would not pay for a $10,000 data feed from Reuters to the back office.

The Risk Assessment analyst discovered that the spreadsheet was corrupt as the cells for the Yen and Euro - the two currencies in which Rusnak traded the most - had links to Rusnak's computer that detoured outside of Reuters.

Fourteen months after the risk assessment analyst discovered that the source of daily foreign exchange rates was not independent, and approximately six months after she discovered that the rate spreadsheet was corrupt, Allfirst finally remedied the problem:

> *"..... investigation by AIB trading experts has revealed that Mr. Rusnak ... seems to have halted the practice (of price manipulation) only in April 2001, around the time risk assessment obtained a copy of the pricing spreadsheet he had corrupted"*

## 3.0 WHAT LESSONS ARE HERE FOR THE AUDIT AND CONTROL OF SPREADSHEETS?

The author has previously stated that:

> *"...an error in a spreadsheet application can subvert all the controls in all of the systems which feed data into it…".[Butler, 2000]*

The Rusnak./ Allfirst case is the perfect illustration of this subversion [Chadwick, 2003]. We must therefore be alert for the possibility of deliberate manipulation of spreadsheets as well as "*innocent*" mistakes. We must ensure that all important spreadsheets we encounter are risk assessed and tested where appropriate. We must ensure that key data items and constants are checked to source and critically reviewed. (The AIB Interim reports suggest that whilst the calculations in the VaR model were correct, Rusnak interfered with the DATA).

The use of spreadsheet error detection methodologies [Ayalew et al, 2000] [Croll, 2002] and spreadsheet auditing tools [Nixon & O'Hara, 2001] such as SpACE would have highlighted the existence of the links in the pricing spreadsheet and explained where they went to enable follow-up. More generally, in order to minimize risks, it is essential that very early on in the audit process clear investigative steps are taken.

**4.0 CONCLUSION**

If potential weaknesses or risks in a system are detected, they must be followed up. The internal audit and risk managers in Allfirst were criticised for failing to follow up indications of risk - sometimes because of a lack of understanding of the business areas concerned and sometimes because they placed too much reliance on the (corrupted) key controls and failed to use the corroborative records available to them. These failures were, it should be noted, symptoms of weak or non-existent management control and corporate governance.

**APPENDIX A**

VaR is a statistical measure used to estimate the maximum range of loss that is likely to be suffered in a given portfolio. Allfirst used a VaR model developed by AIB Group. The specific application used in Allfirst to measure Allfirst VaR and stop-loss was developed in-house by Allfirst treasury risk control personnel.

Using Monte Carlo simulation techniques, the model generates 1,000 hypothetical foreign exchange spot and volatility rates and calculates the resulting profit or loss. The value-at-risk is equated to the tenth-worst outcome produced by the simulation, which yields a 99 percent level of statistical certainty.

Note that the AIB interim report is incorrect as the tenth-worst (ie 10% or 90% decile) outcome would, at best, yield only a 90 per cent level of statistical reliability.

**REFERENCES**

AIB Group, (2002), Report to the Boards of Allied Irish Banks, p.l.c., Allfirst Financial Inc. and Allfirst Bank Concerning Currency Trading Losses, Promontory Financial Group And Wachtell, Lipton, Rosen & Katz, March 12, 2002, published by Allied Irish Banks plc http://www.aibgroup.com (Follow the links to Investor Relations 14 Mar 2002) or see http://www.aibgroup.com/servlet/ContentServer?pagename=AIB_Investor_Relations/AIB_Press_Releas/aib_d_press_releases&c=AIB_Press_Releas&cid=1015597171590&channel=IRCA  Accessed 11[th] October 2009 20:20

Ayalew, Y., Clermont, M., Mittermeir, R. (2000), "Detecting Errors in Spreadsheets", Proc. European Spreadsheet Risks Int. Grp. (EuSpRIG) 2000 51-63 ISBN:1 86166 158 4, http://arxiv.org/abs/0805.1740

Butler, R. J., (2000) "Risk Assessment For Spreadsheet Developments: Choosing Which Models to Audit", Proc. European Spreadsheet Risks Int. Grp. (EuSpRIG) 2000 65-74 ISBN:1 86166 158 4, http://arxiv.org/abs/0805.4236

Chadwick, D., (2003) "Stop the Subversive Spreadsheet!", IFIP Integrity and Internal Control in Information Systems, Vol 124, pp. 205-211, Kluwer, http://aps.arxiv.org/abs/0712.2594

Croll, G.J., (2003), "A Typical Model Audit Approach: Spreadsheet Audit Methodologies in the City of London", IFIP, Integrity and Internal Control in Information Systems, Vol 124, pp. 213-219, Kluwer, 2003, http://aps.arxiv.org/abs/0712.2591

Nixon, D., O'Hara, M., (2001) "Spreadsheet Auditing", Proc. European Spreadsheet Risks Int. Grp. 2001 ISBN:1 86166 179 7